\documentclass[10pt,twocolumn]{IEEEtran} 

\usepackage{fancyhdr}
\usepackage{epsfig}
\usepackage{threeparttable}
\usepackage{epsf,epsfig}
\usepackage{amsthm}
\usepackage{amsmath}
\usepackage{amssymb}
\usepackage{amsfonts}
\usepackage[noadjust]{cite}
\usepackage{dsfont}
\usepackage{subfigure}
\usepackage{color}
\usepackage{enumerate}
\usepackage{comment}

\newtheorem{theorem}{Theorem}

\newtheorem{definition}[theorem]{Definition}

\newtheorem{proposition}{Proposition}

\def\E{\mathsf{E}}
\def\SINR{\mathsf{SINR}}

\def\SIR{\mathsf{SIR}}

\def\({\left(}
\def\){\right)}
\def\lb{\left\{}
\def\rb{\right\}}
\def\[{\left[}
\def\]{\right]}

\newcommand{\nn}{\nonumber}

\includecomment{comment}	

\def\lambdahM{{\hat{\lambda}_m}{}}

\def\lambdahMu{{\hat{\lambda}_\mu}{}}
\def\lambdaM{\lambda_{m}}
\def\lambdaMu{\lambda_\mu}
\def\lambdaU{\lambda_{u}}
\def\gammaM{\gamma_{m}}
\def\gammaMu{\gamma_\mu}
\def\gammaMUL{\gamma_{m.u}}
\def\gammaMDL{\gamma_{m.d}}
\def\gammaMuUL{\gamma_{\mu.u}}
\def\gammaMuDL{\gamma_{\mu.d}}
\def\WM{W_\text{m}{}}
\def\WMu{W_\text{$\mu$}{}}
\def\WMUL{W_\text{m.u}{}}

\def\WMuUL{W_\text{$\mu$.u}{}}

\def\RL{R_{L}{}} 
\def\CP{C_P{}}
\def\CLt{C_{L}^{(t)}}
\def\CL{C_{L}{}}

\def\rhoM{\rho_{m}}

\def\alphaM{\alpha_{\emph{\text{m}}}}
\def\alphaMu{\alpha_\mu}
\def\RDL{\mathcal{R}_\text{d}{}}
\def\RUL{\mathcal{R}_\text{u}{}}

\def\PAPR{\text{PAPR}_\text{u}}
\def\PhiActU{\Phi_{\dot{\text{u}}}}
\def\PhiActM{\Phi_{\dot{\text{m}}}}
\def\PhiActMu{\Phi_{\dot{\mu}}}

\def\papertitle{ Tractable Resource Management in Millimeter-Wave Overlaid Ultra-Dense Cellular Networks }

\IEEEoverridecommandlockouts 

\begin{document}

\title{ \fontsize{19.8}{20}\selectfont \papertitle}

\author{Jihong Park, Seong-Lyun Kim, and Jens Zander
\thanks{J. Park and S.-L. Kim are with Dept. of Electrical \& Electronic Engineering, Yonsei University, Seoul, Korea (email: jhpark.james@yonsei.ac.kr, slkim@yonsei.ac.kr).  }
\thanks{J. Zander is with Wireless@KTH, KTH -- The Royal Institute of Technology, Stockholm, Sweden (email: jensz@kth.se).}}

\maketitle

\begin{abstract} 
\emph{What does millimeter-wave (mmW) seek assistance for from micro-wave ($\mu$W) in a mmW overlaid 5G cellular network?} This paper raises the question of whether to complement downlink (DL) or uplink (UL) transmissions, and concludes that $\mu$W should aid UL more. Such dedication to UL results from the low mmW UL rate due to high peak-to-average power ratio (PAPR) at mobile users. The DL/UL allocations are tractably provided based on a novel closed-form mm-$\mu$W spectral efficiency (SE) derivation via stochastic geometry. The findings explicitly indicate: (i) both DL/UL mmW (or $\mu$W) SEs coincidentally converge on the same value in an ultra-dense cellular network (UDN) and (ii) such a mmW (or $\mu$W) UDN SE is a logarithmic function of BS-to-user density ratio. The corresponding mm-$\mu$W resource management is evaluated by utilizing a three dimensional (3D) blockage model with real geography in Seoul, Korea.
\end{abstract}

\section{Introduction}
The scarcity of micro-wave ($\mu$W) cellular frequency for achieving the 1,000x higher data rate in 5G brings about the overlay of millimeter-wave (mmW) frequency that provides hundreds times more spectrum amount \cite{WRoh:14, Rappaport:14, Ericsson5G:13}. Such an approach however has two major drawbacks. Firstly, mmW signals are vulnerable to physical blockages, yielding severe distance attenuation for both downlink (DL) and uplink (UL) transmissions. Secondly, utilizing extremely wide mmW bandwidth makes UL mmW transmissions at mobiles demanding due to high peak-to-average-ratio (PAPR) \cite{SamsungmmWave:11}. It leads to significant rate difference between DL and UL. This DL/UL rate mismatch may hinder consistent user experiences during DL/UL symmetric services such as video over LTE or cloud gaming. Furthermore, it may even decrease DL rate when the UL rate cannot cope with the required control signals for DL.

For the first blockage problem, the densification of mmW BSs is a straightforward remedy without procuring additional spectrum, leading to equally improve both DL and UL rates. In spite of such a solution, the second DL/UL rate asymmetry problem still remains, of which the DL/UL rate ratio holds the same amount.


To tackle this bottleneck, we focus on the role of the incumbent $\mu$W BS densification as well as its DL/UL resource allocation. Specifically, our prime concern is maximizing the overall (mmW and $\mu$W) DL rate while guaranteeing a minimum UL rate. For given mmW resource amount and BS density, this paper answers the question \emph{how much amount of $\mu$W resource should be allocated to UL transmissions}.

This resource management solution hinges on the DL/UL mmW and $\mu$W spectral efficiencies (SEs). From its analytic perspective, we apply the sate-of-the-art stochastic geometric approaches for the mmW blockage model in \cite{Heath:13, Heath:14,Singh:14}. In addition, we further develop the three-dimensional (3D) blockage model proposed in \cite{Heath:13} based on the actual geography of the buildings in Seoul, Korea. Combined with the 3D mmW blockage model, we derive the maximized overall DL rate in a closed-form on the basis of our preliminary SE analysis in an ultra-dense cellular regime \cite{JHPark:14}. The overall UL rate is also provided in a closed-form by means of a novel upper/lower bound technique. Such closed-form results enable the tractable optimization of the resource management, and thereby provide the design guidelines of the mmW overlaid ultra-dense cellular network (UDN).

\begin{figure}
	\centering
	\includegraphics[width=9cm]{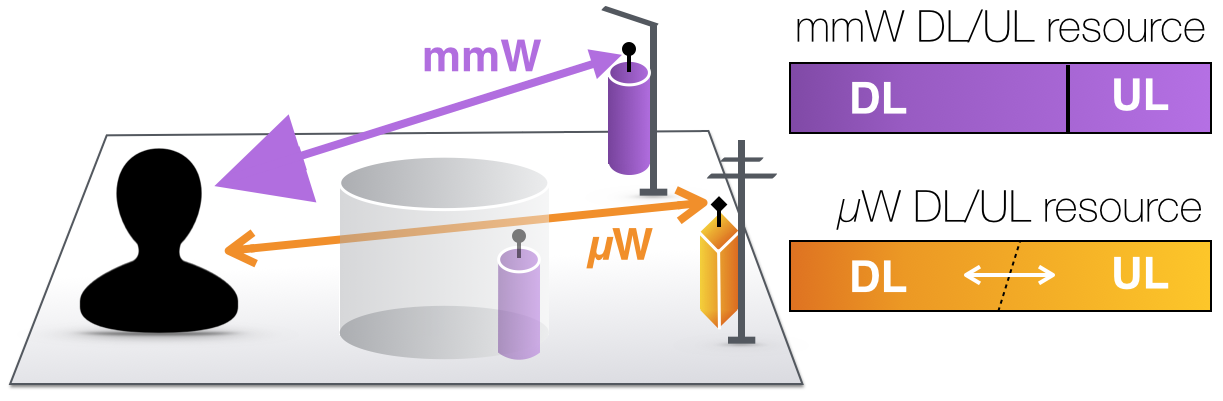}
	\caption{DL/UL resource management and transmission scenario in a mmW-$\mu$W two-tier 5G cellular network. In contrast to $\mu$ transmissions, mmW UL transmissiosn are: (i) blockage non-penetrable and (ii) DL/UL asymmetric since mmW UL resource allocation is restricted by the PAPR at mobiles. To compensate the mismatch, $\mu$W DL/UL resource allocation is optimized in order to maximize the overall DL rate with a minimum UL rate constraint.} \label{Fig:Association}
\end{figure}

The main contributions of this paper are listed as follows.
\begin{enumerate}[1.]
\item Most of the $\mu$W resource should be dedicated to UL transmissions in order to resolve the UL bottleneck in a mmWave overlaid ultra-dense cellular networks: for $500$ MHz mmW bandwidth with $20$\% UL/DL minimum rate ratio requirement, more than $60$\% of the $\mu$W spectrum should be allocated to UL; for $1$ GHz mmW bandwidth, the entire $\mu$W should be dedicated to UL (see Fig. 5).
\item The $\mu$W and mmW SEs for both DL and UL are derived in closed forms that reveal BS densifications logarithmically increase the SEs (see Theorems 1 and 2).
\item Numerical results are verified with a 3D mmW blockage model with the actual building geography in Seoul, Korea (see Figs. 4 and 5 with Table I).
\end{enumerate}


\section{System Model}
\subsection{Millimeter-Wave Overlaid Cellular Network} \label{Sect:Network}
The proposed two-tier network comprises: (i) mmW BSs whose locations follow a two-dimensional (2D) homogeneous Poisson point process (PPP) $\Phi_\text{m}$ with density $\lambdaM$; and (ii) $\mu$W BSs whose coordinates follow a homogeneous PPP $\Phi_\mu$ with density $\lambdaMu$, independent of $\Phi_\text{m}$.

Mobile user coordinates independently follow a homogeneous PPP $\Phi_\text{u}$ with density $\lambdaU$. Without loss of generality, $\Phi_\text{u}$ represents both DL and UL users. For simplicity, we hereafter only focus on outdoor users whose coordinates do not overlap with building blockages. Users simultaneously associate with mmW and $\mu$W BSs via different antennas for each as in macro-and-small-cell dual connectivity \cite{Rel12Beyond:13}. The associations for both mmW and $\mu$W guarantee the maximum received signal powers. The corresponding wireless control signals are communicated via $\mu$W so that it can always guarantee the connections regardless of blockages. The volume of the control signal transmissions are much less than that of data transmissions, and thus is assumed to be negligible.

For multiple associations of users, each BS selects a single active user per unit time slot according to a uniformly random scheduler, resulting in an active user PPP $\PhiActU$. Now that BSs having no associated users are turned off, such a user selection also leads to active mmW and $\mu$W BS PPPs $\Phi_{\dot{\text{m}}}, \Phi_{\dot{\mu}}$ where the BSs associate with at least a single active user.

\subsection{Blockage Model} \label{Sect:SysBlockage}
Consider building blockages that cannot be penetrated by mmW signals. Each single mmW signal reception should therefore ensure line-of-sight (LOS) between the transmitter and receiver pair. Assume that a given transmitter-receiver distance $r$ satisfies the LOS condition if it does not exceed a certain average LOS distance $\RL$. An LOS indicator function $\mathds{1}_{\RL}(r)$ represents such a model, which returns $1$ if LOS exists, i.e. $r \leq \RL$; otherwise $0$. This average LOS distance model proposed by \cite{Heath:13} corresponds with the approximation of a Boolean blockage model \cite{Heath:13,HallBook:CoverageProcess:1988}, and is similar to distance threshold blockage models \cite{Singh:14,Heath:14}. 

For the 3D blockage model, $\RL$ is directly computed by utilizing building perimeters, areas, coverage from actual geographical data, to be specifically elaborated in Section \ref{Sect:3DBlockage}. The value of $\RL$ with the 2D blockage model \cite{Heath:13,Heath:14,Singh:14,Kulkarni:14} neglects building height data during its computation. Throughout this paper, we by default consider the 3D blockage model unless otherwise noted, but apply a 2D channel model that provides more tractability. Incorporating a 3D channel model is deferred to our future work.

In contrast to mmW signals, $\mu$W$\mu$W signals are not affected by blockages thanks to their high diffraction and penetration characteristics. 
 

\subsection{Channel Model}
A mmW antenna array at a BS (or user for UL) directionally transmits a signal with unity power to its associated user (or BS). For simplicity, the antenna array gains at mmW transmitters and receivers are assumed to be unity. The transmitted mmW signal experiences path loss attenuation with the exponent $\alphaM >2$ as well as Rayleigh fading with unity mean, i.e. channel fading power $g \sim \exp(1)$.

The transmitted directional signal has a main lobe beam width $\theta$ radian, and its beam center $\Theta \in [0, 2\pi]$ points at the associated user (or BS). The received signal powers at the same distances within the mainlobe are assumed to be identical, and sidelobes are neglected for simplicity. Accordingly, at a receiver, interferers are the transmitters whose main lobes cover the location of the receiver. When interference is treated as noise, by the aid of Slyvnyak's theorem \cite{StoyanBook:StochasticGeometry:1995}, DL/UL mmW $\SINR$ for a typical outdoor user is respectively represented  as:
\begin{align}
\SINR_{\text{m.d}} &:=  \frac{ g r^{-\alpha_\text{m}} \mathds{1}_{\RL}(r) }{\sum_{i\in \PhiActM} \Theta_i g_i {r_i}^{-\alpha_\text{m}} \mathds{1}_{\RL}(r_i)+ \sigma^2}\\
\SINR_{\text{m.u}} &:=  \frac{ g r^{-\alpha_\text{m}} \mathds{1}_{\RL}(r) }{\sum_{i\in \PhiActU} \Theta_i g_i {r_i}^{-\alpha_\text{m}} \mathds{1}_{\RL}(r_i)+ \sigma^2}
\end{align}
where $r$ and $r_i$ respectively denote the distances to the associated BS (or user) and interfering BSs (users) from the origin, $\sigma^2$ noise power, and $\Theta_i$ the probability that the $i$-th transmitter interferers with the typical receiver, i.e. the transmitter's beam pointing at its serving receiver with the direction $\Theta$ and beam width $\theta$ covers the typical receiver. Note that $\Theta_i$ is a random variable due to its dependency on the locations of the interfering transmitters' associated receivers that are random variables.

A $\mu$W transmitted signal with unity power experiences the same channel model as in mmW signals only except for its different path loss exponent $\alpha_\mu>2$ and neglecting blockages. At a typical outdoor user, the $\mu$W $\SINR$ is then given as:
\begin{align}
\SINR_{\mu.\text{d}} &:= \frac{g r^{-\alpha_\mu} }{\sum_{i\in \PhiActMu} g_i {r_i}^{-\alpha_\mu}+ \sigma^2} \\
\SINR_{\mu.\text{u}} &:= \frac{g r^{-\alpha_\mu} }{\sum_{i\in \PhiActU} g_i {r_i}^{-\alpha_\mu}+ \sigma^2}.
\end{align}

Now that our interest of this paper is confined to ultra-dense cellular networks, we hereafter only consider signal-to-interference-ratio ($\SIR$) instead of $\SINR$ by neglecting $\sigma^2$ unless otherwise described.

\section{Spectral Efficiencies of Millimeter-Wave Overlaid Ultra-Dense Cellular Networks}
This section derives closed-form mm-$\mu$W SEs, defined as ergodic capacity $\E \log[1 + \SINR]$ in units of nats/sec/Hz (1bit $\approx$ 0.693 nats), which is to be utilized for resource allocation and cell planning in Section \ref{Sect:Opt_WandDens}. For the mmW SE, average LOS distance $\RL$ is calculated under a 3D blockage model based on the actual building geography in Seoul, Korea. 

We hereafter let the subscripts $\text{m}$ and $\mu$ denote mmW and $\mu$W respectively. The representations $\alpha$ and $\lambda$ without subscripts can be either mmW or $\mu$W, which improves notational reusability.

For a BS density $\lambda$, let $\hat{\lambda}$ denote $\lambda / \lambdaU$. This ratio of BS density to user density determines the ultra-densification of a network as follows.
\begin{definition}\emph{(UDN) }
A cellular network with BS density $\lambda$ is called a UDN where $\hat{\lambda} \gg 1$. The notation $f\gg g$ implies f is sufficiently large such that $O\(g/f\)$ is approximated by $0$.
\end{definition}

In a UDN regime, both $\mu$W DL SE $\gammaMuDL$ and UL SE $\gammaMuUL$ share the same upper and lower bounds provided in the following proposition.

\begin{proposition}\emph{($\mu$W DL/UL UDN SE Bounds)}
In a UDN regime, $\mu$W DL SE \emph{$\gammaMuDL$} and UL SE \emph{$\gammaMuUL$} are identically bounded as:
\emph{\begin{align}\label{Eq:Bounds_DLMu}
\gammaMuDL \text{ (or $\gammaMuUL$)} &\geq \log\( 1 + \[ {\rho_{\mu}}^{-1} \lambdahMu \]^{{\frac{\alphaMu}{2}}}\)  -\frac{\alphaMu}{2}; \nn\\
&\leq  \log\( 1 + \[ \( 1 + \frac{2}{\alphaMu}\)\lambdahMu\]^{\frac{\alphaMu}{2}}\) - \frac{\alphaMu}{2}
\end{align}}
where $\rho_{\mu} :=  \int_0^\infty \frac{du}{1 + u^{\frac{\alphaMu}{2}} } = \(\frac{2 \pi}{\alphaMu}\)\text{csc}\(\frac{2 \pi}{\alphaMu}\)$.\\
\emph{\begin{proofsk} 
Let $\rho_\mu^{(t)}$ denote $\int_{(e^t-1)^{-\frac{2}{\alpha_\mu}}}^\infty  \frac{du}{1 + u^{\frac{\alpha_\mu}{2}}} $. The representation of $\gammaMuDL$ (or $\gammaMuUL$) is given as follows via rephrasing the equation (16) in \cite{Andrews:2011bg}:
\begin{align}
\int_{t>0} \E_R \[  \exp\( - \lambdaU \pi  r^2 (e^t - 1)^{\frac{2}{\alpha} }  \rho_\mu^{(t)}  \) \] dt \label{Eq:PfProp1ExactPre}
\end{align}
where $R$ denotes the distance to the BS-to-user association distance. Taylor expansion approximates the integrand in \eqref{Eq:PfProp1ExactPre} since $\hat{\lambda} \gg1$ by Definition 1, i.e.
\begin{align}
\E_{R}\[ e^{ -\lambdaU \pi r^2 (e^t - 1)^{\frac{2}{\alpha} }  \rho_\mu^{(t)} }\] &\approx \[ 1 - \hat{\lambda}{}^{-1} (e^t - 1)^{\frac{2}{\alpha} }  \rho_\mu^{(t)}  \]^+. \label{Eq:Taylor}
\end{align}
Consider the upper and lower bounds of $\rho_\mu^{(t)}$. Increasing the integration range trivially yields the upper bound. Applying Taylor expansion with a minor modification leads to the lower bound. The bounds is then provided as below.
\begin{align}
\(1 + \frac{2}{\alpha_\mu} \)^{-1}-(e^t-1)^{-\frac{2}{\alpha_\mu}}\leq \rho_\mu^{(t)} \leq \rho_\mu \label{Eq:rhobounds}
\end{align}
Plugging \eqref{Eq:Taylor} and \eqref{Eq:rhobounds} into \eqref{Eq:PfProp1ExactPre} results in the desired result.
\hfill $\blacksquare$ 
\end{proofsk}}
\end{proposition}

As $\mu$W BS density increases, the upper and lower bounds of the DL/UL SEs tend to converge on the identical closed-form $\mu$W UDN SE presented in the following theorem.

\setcounter{theorem}{0}
\begin{theorem}\emph{(Asymptotic $\mu$W UDN SE)  }
For $\lambdahMu \rightarrow \infty$, DL and UL $\mu$W SEs identically converge to $\gammaMu$, i.e.
\emph{$\gammaMu = \lim\limits_{\lambdahMu \rightarrow \infty}\gammaMuUL = \lim\limits_{\hat{\lambda}_\mu \rightarrow \infty}\gammaMuDL$}, which is given as below.
\begin{eqnarray} 
&&\hspace{-20pt} \gammaMu = \frac{\alphaMu}{2} \log \lambdahMu \hfill 
\end{eqnarray}
\end{theorem}
The results above explicitly reveal that $\mu$W DL/UL UDN SEs logarithmically increase with their BS-to-user density ratios. The SEs also increase with path loss exponents, which is in accordance with the well-known behavior under an interference-limited regime \cite{Andrews:2011bg}.

Similarly, the upper and lower bounds of DL mmW SE $\gammaMDL$ and UL SE $\gammaMUL$ are given in the following proposition.

\begin{proposition}\emph{(mmW DL/UL UDN SE Bounds)}
In a UDN regime, mmW DL SE \emph{$\gammaMDL$} and UL SE $\gammaMUL$ are identically bounded as:
\emph{
\small\begin{align} 
\gammaMDL\text{ (or $\gammaMUL$)}  &\geq \int_{t>0} \CLt \( 1 - \rhoM \lambdahM^{-1} \[ \frac{\theta}{2\pi}(e^t-1)\]^{\frac{2}{\alphaM}}\)^+ dt; \nn\\
&\hspace{-45pt} \leq \int_{t>0} \CLt \( 1- \[\(1 + \frac{2}{\alphaM}\) \lambdahM\]^{-1} \[\frac{\theta}{2\pi}\(e^t - 1\) \]^{\frac{2}{\alphaM}} \)^+ dt  
\end{align}\normalsize}
where $\CLt:=  1 -  e^{ -\lambdaM\pi \RL^2 \lb 1 + \lambdahM^{-1}\rhoM \[ \frac{\theta}{2\pi}(e^t-1)\]^{\frac{2}{\alphaM}}   \rb }$
and $\rhoM :=  \int_0^\infty \frac{du}{1 + u^{\frac{\alphaM}{2}} } = \(\frac{2 \pi}{\alphaM}\)\text{csc}\(\frac{2 \pi}{\alphaM}\)$.\\
\emph{\begin{proofsk}
Let $\rho_m^{(t)}$ denote $\int_{ (e^t-1)^{-\frac{2}{\alphaM}}}^{\(\frac{\RL}{r}\)^2 (e^t-1)^{-\frac{2}{\alphaM}}}  \frac{du}{1 + u^{\frac{\alphaM}{2}}}$. It is bounded as follows.
\begin{align}
1 -\(e^t-1\)^{-\frac{2}{\alphaM}} - \(1 + \frac{\alphaM}{2}\)^{-1}  \leq \rho_m^{(t)}\leq \rho_m
\end{align}
The remainder of the proof follows from the procedure in the proof of Proposition 1 with minor modification.
\hfill $\blacksquare$\end{proofsk}}
\end{proposition}

The mmW DL/UL SEs also asymptotically converge on the identical closed-form mmW UDN SE as stated in the following theorem.

\begin{comment}
\begin{figure}
\centering \hspace{-.2cm}
 	\subfigure[Gangnam (2$\times$2 km$^{2}$)]{\includegraphics[width=2.9cm]{Figs_M/Map_Gangnam.png}   }  \hspace{-.2cm}
 	\subfigure[Jongro (1$\times$1 km$^{2}$)]{\includegraphics[width=2.8cm]{Figs_M/Map_Jongro.png}    }\hspace{-.2cm}
	\subfigure[Yonsei (2$\times$2 km$^{2}$)]{\includegraphics[width=2.7cm]{Figs_M/Map_Yonsei.png}    }
	\caption{Target areas for building statistics: Gangnam, Jongro, and Yonsei, Seoul, Korea.}
\end{figure}
\end{comment}

\begin{table}[!t]
\renewcommand{\arraystretch}{1.1}
\caption{Average LOS Distances Measured from Building Statistics in Seoul, Korea}
\label{Table:GeoStat}
\centering
\begin{tabular}{l | r | r | r }
\hline
 &\textbf{Gangnam} & \textbf{Jongro}& \textbf{Yonsei}\\
Building parameters (unit) &(2$\times$2 km$^2$)&(1$\times$1 km$^2$)&(2$\times$2 km$^2$)\\
\hline\hline
Avgerage perimeter $\rho$ (m$^2$) &59.02&39.29&51.99\\
Avgerage area $A$ (m$^2$) &218.60&107.67&173.95\\
Coverage $\kappa$ (\%) &34.77&46.90&25.48\\
Avgerage height $\E H$ (m) &14.23 & 8.12 & 11.14 \\
Height $\sim$ \text{log$\mathcal{N}$}$(\mu$,$\sigma)$ & (1.62, 0.27) & (0.69, 0.55)& (1.10, 0.34)\\
\hline
2D blockage parameter $\beta$ & 0.073 & 0.014 & 0.056 \\
3D height parameter $\eta$ & 0.36 & 0.22 & 0.13 \\
\textbf{3D LOS distance $R_{\text{L}}$} (m) & \bf{49.61} & \bf{33.33} & \bf198.76 \\
\hline
\end{tabular}
\end{table}

\setcounter{theorem}{1}
\begin{theorem}\emph{(Asymptotic mmW UDN SE)} For $\lambdahM \rightarrow \infty$, DL and UL mmW SEs identically converge to $\gammaM$, i.e.
$\gammaM = \lim\limits_{\lambdahM \rightarrow \infty}\gammaMUL = \lim\limits_{\lambdahM \rightarrow \infty}\gammaMDL$, which is given as
\emph{\begin{align}
\gammaM &= {{\frac{\alphaM \CL}{2}}}   \log\lambdahM 
\end{align}}
where \emph{$\CL := 1 - \exp\(-\lambdaM \pi \RL^2 \)$}.
\end{theorem}

This indicates mmW DL/UL SE logarithmically increases with BS-to-user density ratio as in the $\mu$W SE. The only difference is its average LOS probability $\CL$ due to the blockage vulnerability of mmW signals. The calculation of $\CL$ relies on the average LOS distance $\RL$ of which the derivation is provided in Section \ref{Sect:3DBlockage}.

\begin{figure}
\centering
  \includegraphics[width=9cm]{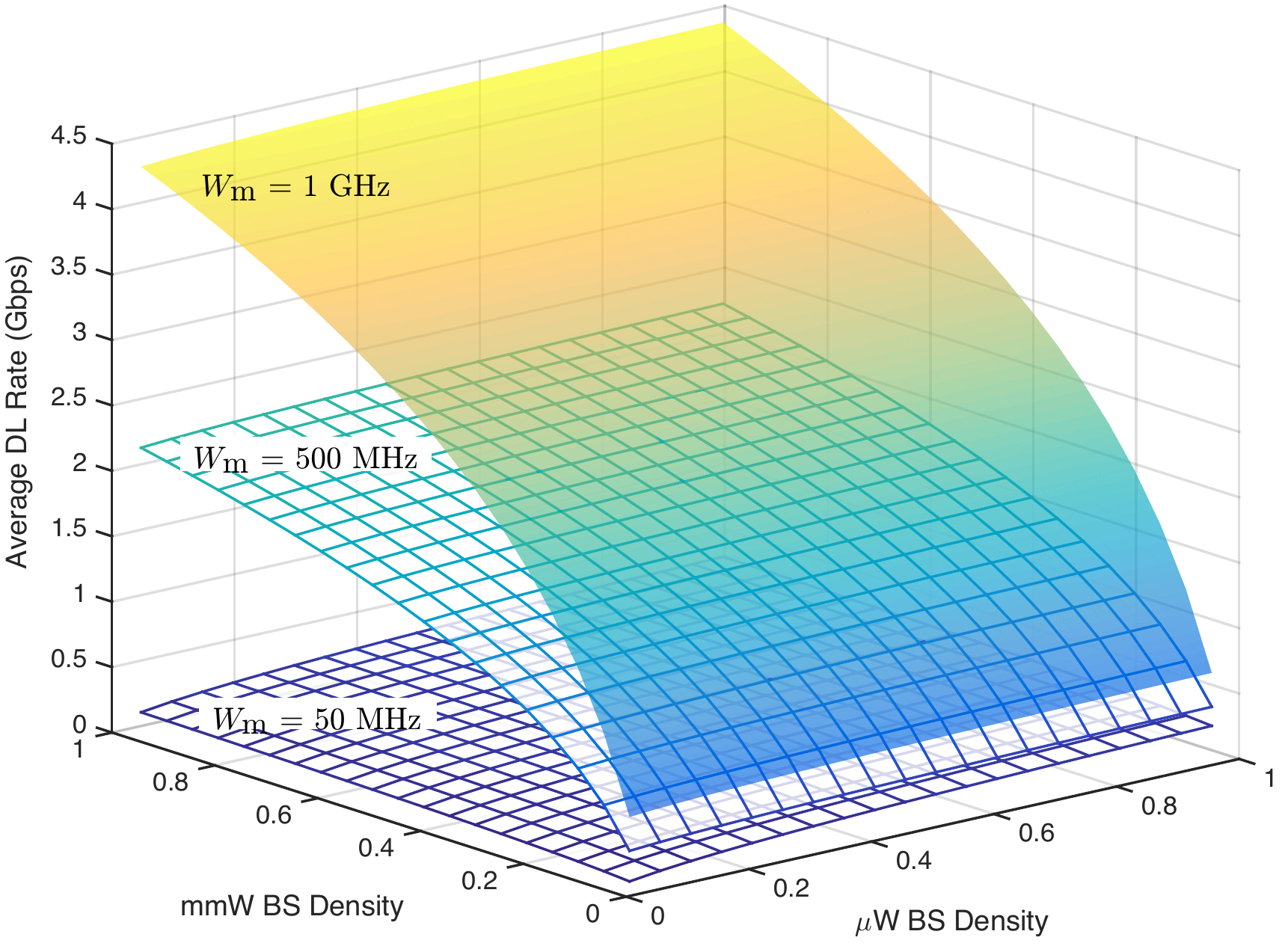}
\caption{Maximized DL average rate ($\WMu = 20$ MHz, $\zeta = 0.2$, $\RL = 100$ m, $f_s = 244.14$ kHz, $\delta = 7$ dB, $\epsilon = 0.7$).} \label{Fig:DLRate3D}
\end{figure}

\subsection{Average LOS Distance under a 3D Blockage Model} \label{Sect:3DBlockage}
This section calculates average LOS distance $\RL$ under a 3D blockage model based on the actual building geography in Seoul, Korea. 

The possibility that a given link with distance $r$ is LOS not only depends on blockage locations but also their heights. We consider both effects in a 3D blockage model proposed in \cite{Heath:13}, and its corresponding 3D LOS distance is given as:

\begin{equation}
\RL = \frac{2(1-\kappa)}{\beta \eta }
\end{equation}
where 
\begin{eqnarray}
\beta &:=& \frac{- 2 \rho \log\( 1 - \kappa\)}{\pi A},\\
\eta &:=& \int_{0}^{1} \Pr\( H \leq (1-s) B\)ds, \label{Eq:BlockageHeight}
\end{eqnarray}
$\rho$ average blockage perimeter, $\kappa$ average building coverage, $A$ average building area, $H$ building height, and $B$ BS height.

Note that $\beta$ and $\eta$ can be calculated via geographical data in practice. By the aid of the ministry of land, infrastructure, and Transport of Korea, we derive such parameters  corresponding to the following three hotsopt regions in Seoul, Korea: Gangnam, Jongro, and Shinchon (Yonsei). The results are summerized in Table \ref{Table:GeoStat}.

The calculations of $\beta$ rely solely on QGIS, an open source geographic information system (GIS) \cite{QGIS}. The process for $\eta$ (or $\RL$) additionally requires the distribution of building height $H$ due to \eqref{Eq:BlockageHeight}. The geographic data does not include the building height information. We detour this problem via the information on the number of building floors. We assume unit floor height is $3$ m, and then derive the building floor distribution via the data curve fitting. Log-normal distributions are well fitted, of which the maximum mean squared error is less than $0.016$. For BS heights $B$, also required in \eqref{Eq:BlockageHeight}, we assume each BS height follows the average building height $\E H$.

\begin{figure}
	\centering
	\includegraphics[width=9cm]{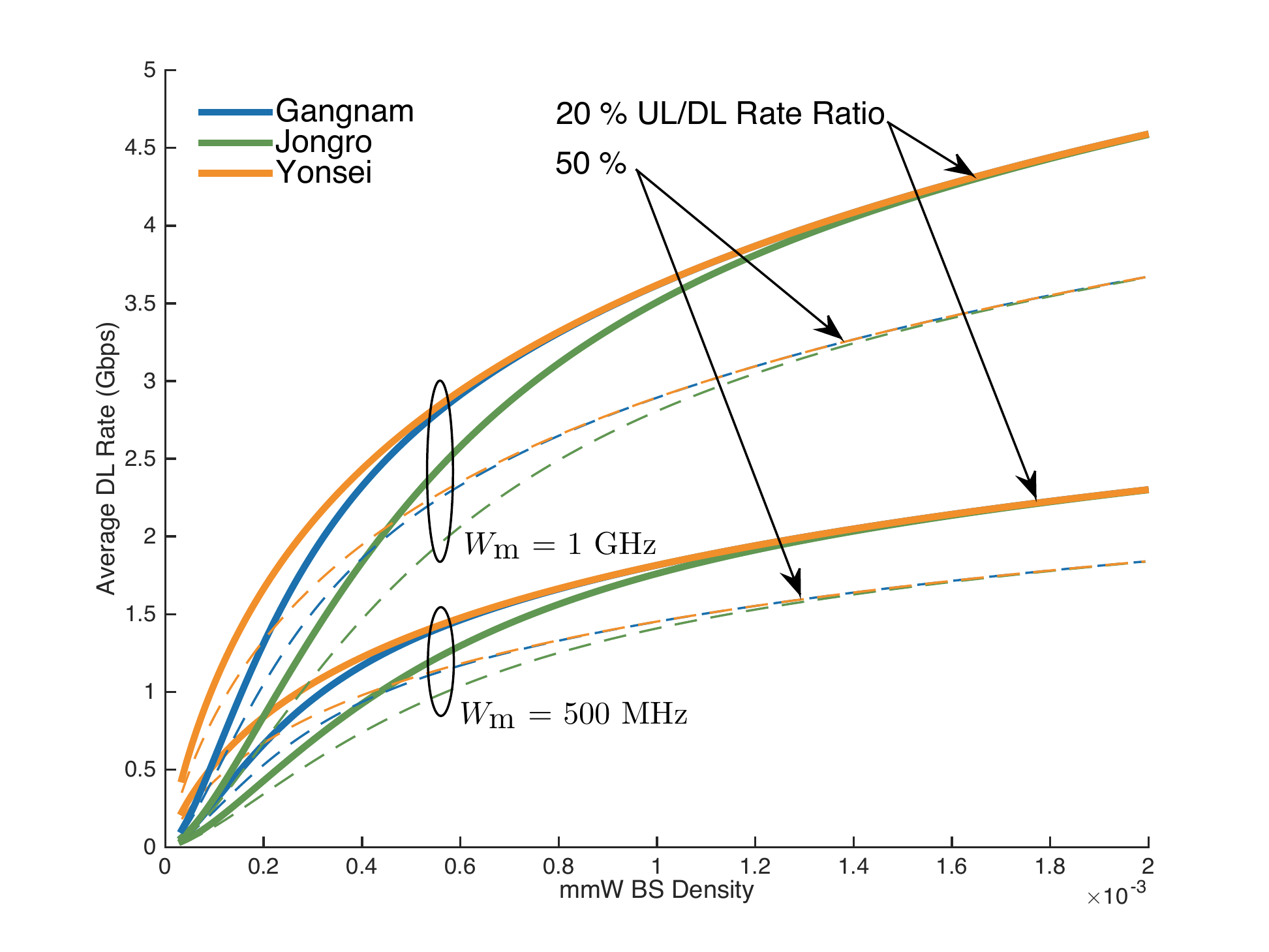}
	\caption{Maximized DL Average rate for different UL/DL rate ratio requirements with actual building geography in Seoul, Korea ($\WMu = 20$ MHz, $\lambdaU = 10^{-4}$, $f_s = 244.14$ kHz, $\delta = 7$ dB, $\epsilon = 0.7$).} \label{Fig:DL_Rate_Zeta}
\end{figure}

\section{Resource Management and Cell Planning in Millimeter-Wave Overlaid Ultra-Dense Cellular Networks}	\label{Sect:Opt_WandDens}
\subsection{DL Average Rate with Minimum UL Rate Constraint}
We consider frequency-division duplex (FDD) $\mu$W and mmW resource allocations with orthogonal frequency-division multiple access (OFDMA) of mobile users. The $\mu$W DL and UL frequency amounts respectively are $W_{\mu.\text{d}}$ and $W_{\mu.\text{u}}$ out of the entire amount $W$ where $W_{\mu.\text{d}} + W_{\mu.\text{u}} = \WMu$.

The mmW UL spectrum amount is limited by $\WMUL$ since increasing the amount incurs PAPR increment at mobile users, denoted as $\text{PAPR}_\text{u}$. The amount is determined such that the PAPR outage is no larger than a target rate $\epsilon$, i.e. $\Pr\( \PAPR> \delta \) \leq \epsilon$ for constants $\delta, \epsilon>0$. According to \cite{Jiang:08}, the $\PAPR$ outage probability is given as
\begin{align}
\Pr\( \PAPR> \delta \) &\approx 1 - \exp\( - \frac{\WMUL e^{-\delta}}{f_s} \sqrt{\frac{\pi \delta}{3}} \)
\end{align}
where $f_s$ denotes the subcarrier spacing. 

Note that the said UL PAPR bottleneck can be relaxed via single-carrier frequency-division multiple access (SC-FDMA). Increasing UL mmW spectrum amount in such a scenario, however, still brings about another bottleneck, for instance, sampling rate overhead at analog-to-digital converter (ADC), of which the modeling is deferred to future work.

Define DL and UL average rates $\RUL$ and $\RDL$ as follows.
\begin{eqnarray}
	\RUL &:=& \WMuUL \gammaMuUL + \WMUL \gammaMUL\\
	\RDL &:=& \( \WMu-\WMuUL \) \gammaMuDL + \( \WM- \WMUL \) \gammaMDL \label{Eq:RateDL} 
	\end{eqnarray}

\begin{figure*}
\centering
	\subfigure[for different mmW resource amount ($\zeta = 0.2$)]{\includegraphics[width=9cm]{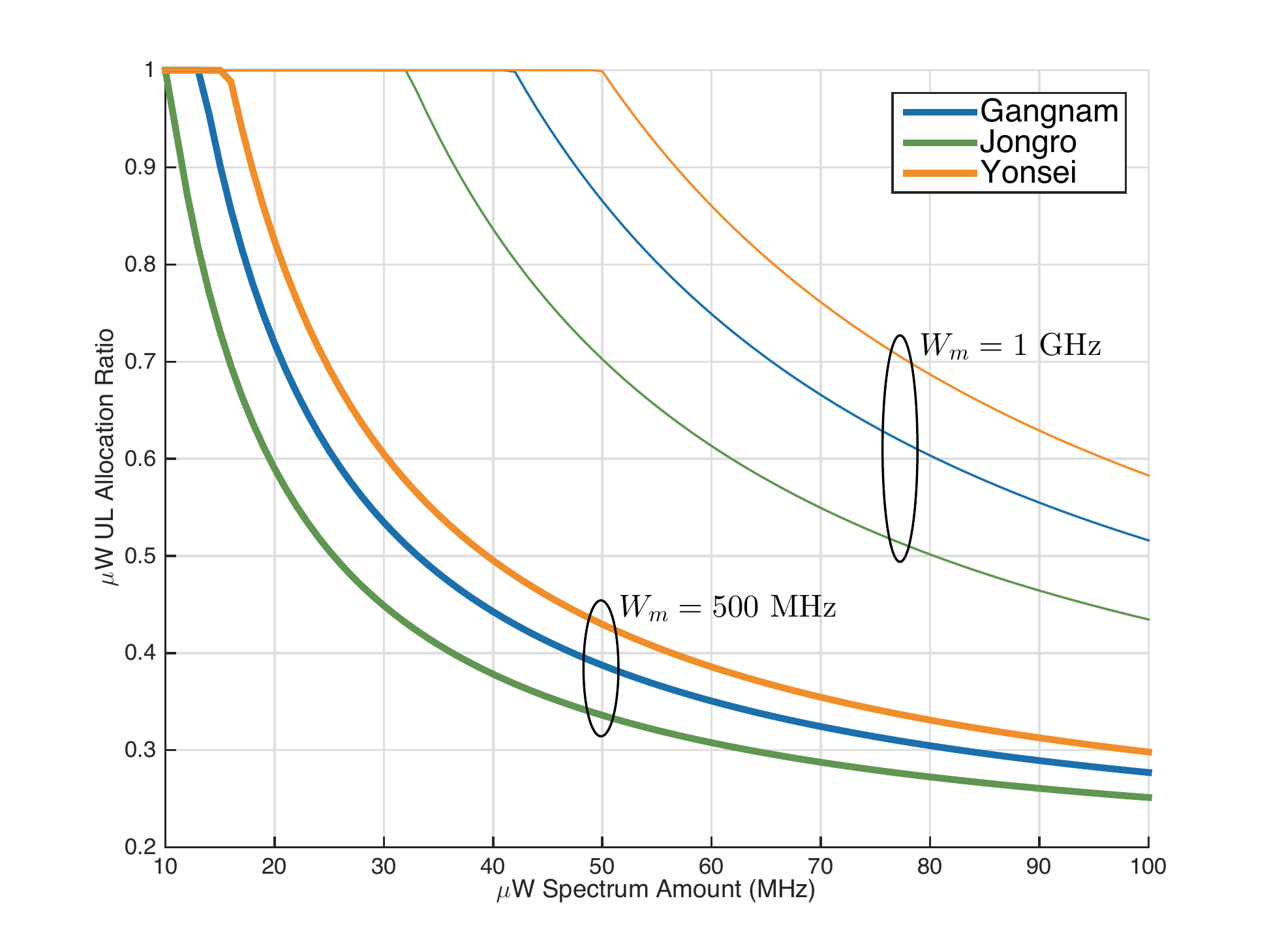}} 	
	\subfigure[for different DL/UL rate ratio requirements ($\WM = 500$ MHz)]{\includegraphics[width=9cm]{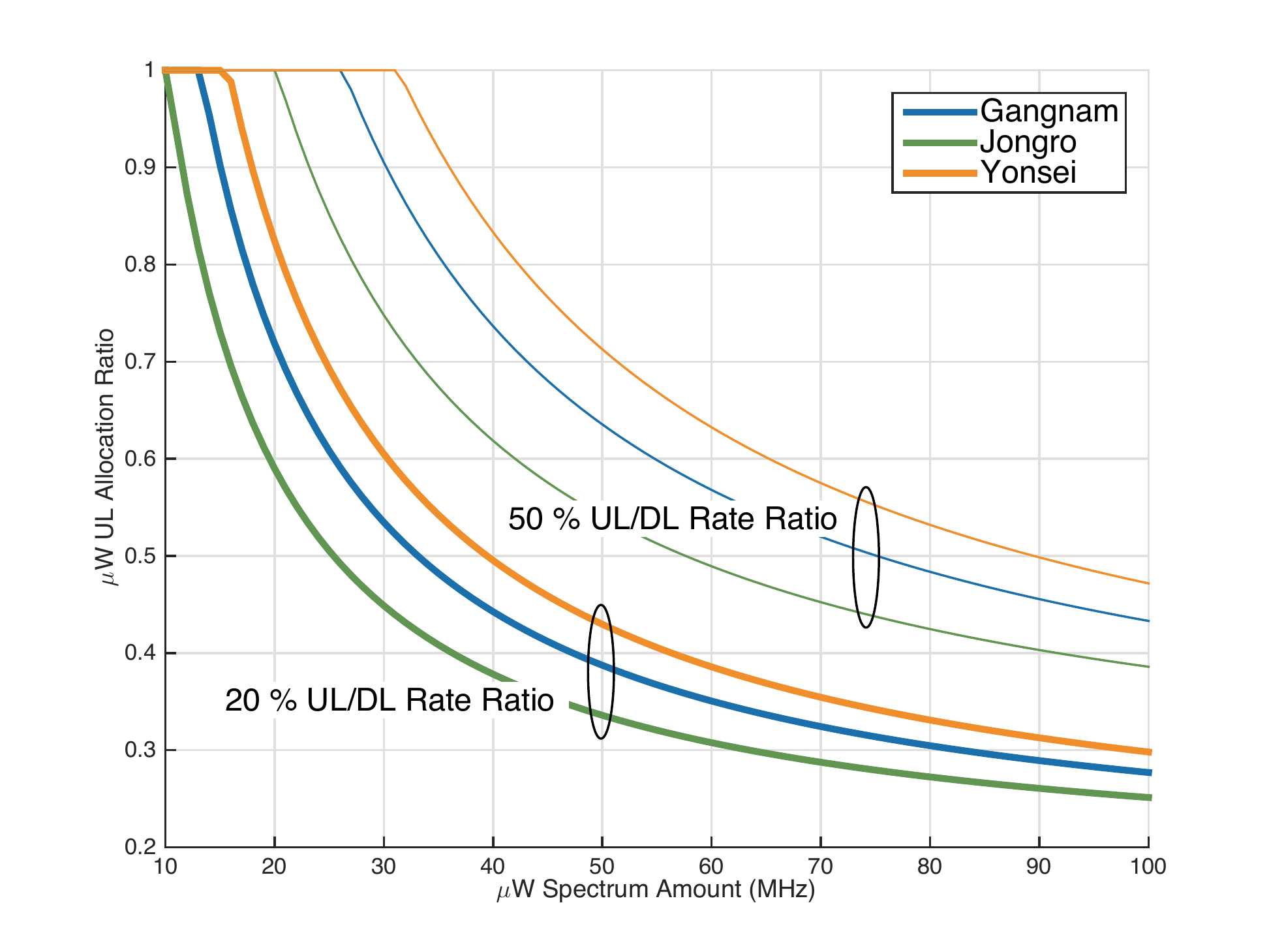}}
	\caption{$\mu$W UL/DL resource allocation for (a) different mmW resource amount and (b) UL/DL rate ratio requirements ($\WMu = 20$ MHz, $\lambdaU = 10^{-4}$, $\lambdaM = 2 \lambdaU$, $\lambdaMu = 4 \lambdaU$, $f_s = 244.14$ kHz, $\delta = 7$ dB, $\epsilon = 0.7$).} \label{Fig:OptW}
\end{figure*}

Let $\zeta \leq 1$ denote the minimum ratio of UL to DL rates. We consider the following problem:
\begin{align}
      \textsf{P1}.&
   \begin{aligned}[t]
    & \underset{ \WMuUL  }{\text{maximize}} \; \RDL  \notag\\
   \end{aligned}   \notag\\
   &\text{subject to} \notag \\
   & \quad   \RUL/\RDL   \geq  \zeta \label{Eq:ULQoS} \\
   &\quad  \Pr\( \PAPR> \delta \) \leq \epsilon. \label{Eq:ULPAPR}
\end{align} 

The objective function is maximized when the equalities in \eqref{Eq:ULQoS} and \eqref{Eq:ULPAPR} hold, leading to the following $\mu$W UL allocation.

\begin{proposition} (Max. DL UDN Rate) \emph{
For $\lambdahM, \lambdahMu \rightarrow \infty$, the maxmum DL average rate with minimum UL constraint is given as follows.
\begin{equation} 
\RDL^{*} = \frac{1}{2(1 + \zeta)} \log\( \lambdahM^{\alphaM \CL  \WM} \lambdahMu^{\alphaMu \WMu} \)
\end{equation} 
}\end{proposition}

The result indicates the overall DL rate increases with the entire $\mu$W and mmW spectrum amounts as well as logarithmically with the BS densities, illustrated in Fig. \ref{Fig:DLRate3D}. 

Fig. \ref{Fig:DL_Rate_Zeta} visualizes the overall DL rate with the actual geographical information listed in Table I. It shows increasing the minimum UL/DL ratio and/or blockages straightforwardly decreases the DL rate. By utilizing this result, we derive the DL/DL $\mu$W resource allocation to cope with the minimum UL rate bottleneck in the following section.

\subsection{Resource Management} \label{Sect:Opt_WandDens2}
This section provides the DL/DL $\mu$W resource allocation for DL average rate maximization with the minimum UL rate constraint. When the equalities in \eqref{Eq:ULQoS} and \eqref{Eq:ULPAPR} hold, putting them into the DL rate definition in \eqref{Eq:RateDL} provides the maximized DL rate. Exploiting the $\mu$W and mmW DL/UL SEs respectively derived in Theorems 1 and 2 then yields the following proposition.

\begin{proposition} (Optimal $\mu$W UL Resource Allocation) \emph{
For $\lambdahM, \lambdahMu \rightarrow \infty$, DL average rate maximizing $\mu$W resource allocation under a minimum UL constraint is given as below.
\begin{equation} 
\WMuUL^* = 
\frac{\WMu}{1 + \zeta^{-1}} + \(\frac{\WM}{1 + \zeta^{-1}} - \CP \) \frac{\alphaM \CL \log \lambdahM}{\alphaMu \log \lambdahMu}
\end{equation}  
where $\CP := \sqrt{3} f_{s} e^{{\delta}} \(  \pi\delta \)^{-\frac{1}{2}}\( \log\epsilon^{{-1}}\)^{{-1}}$
}\end{proposition}

The implications of this result are visualized in Fig. \ref{Fig:OptW}. Even when the minumum UL/DL rate is only $20$\% at $20$ MHz $\mu$W bandwidth, the UL $\mu$W spectrum allocation becomes larger than the half. In addition, the $\mu$W UL allocation require more spectrum as mmW spectrum amount increases, as shown in Fig. 5(a). Such an allocation is in the opposite direction to the current resource management trend seeking more DL resources. 

For much higher UL rate requirement in accordance with the forthcoming proliferation of UL consuming services such as video over LTE and cloud gaming, the entire $\mu$W spectrum should be dedicated to UL transmissions, as illustrated in Fig. 5(b). This partitions the roles of mmW and $\mu$W respectively devoted for DL and UL, opening an interesting resource management direction for 5G.

\section{Conclusion}
This paper investigates the mmW's innate UL rate bottleneck in a mmW overlaid 5G UDN, and provides a solution from the perspective of the resource management and cell planning in a tractable manner. Such a intuitive result was viable owing to the novel closed-form UDN SEs (see Theorems 1 and 2).

The result implies that the incumbent $\mu$W should aid UL more to improve the overall DL rate by resolving the UL rate bottleneck. For larger mmW spectrum and/or higher minimum UL rate requirement, $\mu$W may be necessary to be dedicated to UL-only operations while the mmW is responsible for DL transmissions (see Fig. 5). Such results contradicting to the current trend open further issues on resource management. Our tractable results will shed light on resolving these problems.


\bibliographystyle{ieeetr}

\end{document}